\documentclass[a4paper]{jpconf_wPageNos}
\usepackage{graphicx}
\usepackage{amsmath}
\usepackage{braket}
\usepackage{physics}
\usepackage{bm}
\usepackage{pst-all} 
\usepackage{cite}
\usepackage{subcaption}
\usepackage{amssymb}
\newcommand*\diff{\mathop{}\!\mathrm{d}}

\DeclareMathOperator{\sinc}{sinc}

\begin{document}

\title{Superoscillations: Realisation of quantum weak values}

\author{Marc Nairn\footnotemark}

\begin{abstract}
Superoscillations are band-limited functions with the peculiar characteristic that they can oscillate with a frequency arbitrarily faster than their fastest Fourier component. First anticipated in different contexts, such as optics or radar physics, superoscillations have recently garnered renewed interest after more modern studies have successfully linked their properties to a novel quantum measurement theory, the weak value scheme. Under this framework, superoscillations have quickly developed into a fruitful area of mathematical study whose applications have evolved from the theoretical to the practical world. Their mathematical understanding, though still incomplete, recognises such oscillations will only arise in regions where the function is extremely small, establishing an inherent limitation to their applicability.
This paper aims to provide a detailed look into the current state of research, both theoretical and practical, on the topic of superoscillations, as well as introducing the two-state vector formalism under which the weak value scheme may be realised.
\end{abstract}
\footnotetext{\textit{Department of Physics and Astronomy, University of Glasgow, UK.}  \\ Electronic address: \textbf{\texttt{2329589n@student.gla.ac.uk}}}
\section{Introduction}
In 1988, Aharonov et al. \cite{Aharonovspin} laid out the foundations for what they claimed to be a new ``type" of measurement, one from which seemingly paradoxical results could be obtained\cite{aharonov1847can}. This measurement, under certain conditions, would directly define a new quantum variable, the ``weak value" of a system, which, as opposed to the discrete eigenvalues of a standard quantum system, is unbounded and can yield results beyond its original eigenvalue spectrum.

Behind this peculiar phenomena the mathematics seemed to describe a common, but counter-intuituve, property: band-limited functions locally oscillating faster than their fastest Fourier component\cite{berry1994faster,Berry_1994billiards}, aptly named, superoscillations. Further, these oscillations can be arbitrarily large over arbitrarily long intervals~\cite{Berry_2006_time}. As an extreme example for this, in one of his early papers on the topic \cite{berry1994faster}, Berry devised a way in which superoscillations could be used to reproduce, exactly, Beethoven's ninth symphony in a 1Hz band-limited signal. 

The ``weakness" of this observation comes from the superoscillatory regions corresponding to parts of the function where it is comparatively small. It is, in fact, a requirement for these functions to grow to enormous values for their superoscillations to be significant. 

Despite the term being born in the context of quantum mechanics, the same type of behaviour had been studied for some time in different areas of physics. In signal processing, as pointed out in \cite{berry1994faster}, oversampling (sampling beyond the Nyquist-Shannon limit \cite{marks2012SamplingIntro}) a signal leads to highly unstable regions outside of the sampled regime. More importantly, and it will be a focus of this review, such functions show exciting potential applications in the world of optics.

This phenomenon was first envisioned in 1952 from the perspective of ``superdirectivity" by Toraldo di Francia \cite{Toraldo1952}, making a connection with the microwave and radar research of the time. He believed in the possibility that a lens with arbitrarily small focal spot would enable resolution beyond the conventional Abbe diffraction limit\cite[p.390]{lipson2010optical}.  In \cite{berry1994faster} more powerful forms of microscopy are theorised where, with the aid of a two-variable superoscillatory function, sub-wavelength structures could be imaged. Similarly, Berry also points out how superoscillations may arise naturally in any random function over arbitrarily long intervals. The first experimental demonstration for the optical capabilities of superoscillations is highlighted in \cite{huang2007nanohole} where sub-wavelength ``hot-spots" are observed when focusing light through nanohole quasiperiodic arrays without a contribution from the evanescent field. Further, in \cite{huang2007superresolution} they succeed at using superoscillations as a far-field imaging technique, as theorised by Berry and Popescu \cite{Berry_2006_time} when they established superoscillations could propagate sub-wavelength detail further into the field than what is achievable with evanescent waves. 

Not only do these random waves contain superoscillations, they are also quite common: on average, 1/3 of the area fraction of one such wave will be superoscillatory \cite{dennis2008superoscillation}. More generally, and to address Aharonov's paradoxical results in \cite{Aharonovspin}, for any given superoscillatory function one can find arbitrary discrete eigenvalues beyond its bounded spectrum to have computable, non-zero probabilities \cite{berry2010superweak}. Precisely, in \cite{berry2011spin} it is shown Aharonov's predicted result is unlikely, but not impossible, with the probability of the spin of a spin-half particle being 100,  $P(\overrightarrow{S}=100\hbar)=\frac{1}{120000}$.

The practicality of these superoscillations is rooted in the ability to construct them in a controlled manner. It must, again, be highlighted  superoscillations come at a considerable cost: most of their energy will be concentrated outside the superoscillatory region, as the function grows to enormous values away from it. For optics, this means superoscillatory focus can only be achieved within the superoscillatory region, with most of the energy located in very large, diffuse sidebands. Because of this highly sensitive behaviour, quantifying superoscillations becomes important for their optical use. To do so, various authors have investigated properties like the size of the central spot, field of view (distance from central spot to first sideband) or a measure of energy efficiency, with a brief summary of results reproduced in \cite{rogers2020review}. 

Berry \cite{berry1994faster} argued the ``cost" of superoscillations to be that the function is exponentially smaller in the superoscillatory region compared to its normal oscillations, with the exponent increasing with the size of the interval of superoscillations. 

These relations have now been further fine tuned, with \cite{Kempf2006Nyquist} reporting exponential increase in sideband energy with the number of superoscillations but only polynomial increase with the ``speed" of the signal. The estimate on the degree of this polynomial is found to be 4 in a subsequent paper \cite{Katzav2013Yield}. The exponential relationship is quite prohibitive in terms of potential practical uses for superoscillatory functions, but the polynomial one poses smaller challenges. For optics, as pointed out in \cite{rogers2020review}, if the focused energy decreased exponentially as the spot shrank, then vastly higher laser powers would be needed to realise small spots.

To enable fuller quantitative analysis of superoscillations, a precise definition for them is required. Superoscillations correspond to regions where the local wavenumber, $k$, is larger than the largest frequency component of the wave, $k_\text{max}$. A natural way to measure the rate of change of its oscillations, especially when dealing with more complicated waves, is the phase gradient of the wave \cite{berry2008natural, dennis2008superoscillation}. As such, for a wave of the form:
\begin{equation}
    \psi(\bm{r})=\rho(\bm{r})\exp(\text{i}\phi{\bm{r}})
\label{monochromatic wave}
\end{equation} with wavenumber $k_0$, superoscillatory regions correspond to:\begin{equation}
    k(\bm{r})\equiv\nabla\phi(\bm {r})\;>\;k_0
    \label{phase gradient}
\end{equation}

The connection of superoscillations with the phase gradient arises from early research on phase singularities, understood to be stable points on any wave that can exhibit interesting sub-wavelength phase topology  \cite{berry2019roadmap}. Going around one such point, the phase changes by 2$\pi$, so, as one approaches it, the phase gradient can be arbitrarily large: larger, in fact, than any of its bounded Fourier components. Therefore any wave, especially monochromatic ones \cite{dennis2008superoscillation}, will be superoscillatory near its phase singularities. 

As mentioned, an alternative definition exploits the weak value scheme introduced by Aharonov  involving an operator with a bounded eigenvalue spectrum\cite{Aharonovspin}. Pre and post-selected states then lead then to the weak value of the operator. As is the case for optics, the weak value of the momentum operator (or wavenumber) is explicitly defined as the local wavenumber when the pre-selected state is position and the post-selected state is the wavefield. Therefore, the theoretical framework described above allows for better understanding of the mathematics behind superoscillatory functions. 

This paper will begin by outlying the mathematical and physical reasoning behind Aharonov's weak value scheme with the two-state vector formulation (TSVF) at the center of it. This will serve as motivation to explain the unexpected results one may find in quantum systems and in the context of rapidly oscillating band-limited functions.

It will then introduce some examples of these superoscillations, how they may be constructed or naturally arise, and their peculiar properties. Superoscillations and superoscillatory regions will be quantified with the help of the local wave-vector, alternatively defined as the phase gradient of a complex function. A discussion on the energy of these functions will lead to the conclusion only a small fraction of the total energy is expressed as superoscillations, with very large sidebands concentrating most of it.

The latter sections will focus on making an explicit connection between weak values and superoscillating functions with the aid of pointer shifts in weak measurements, as well as describing the practical use of optimised superoscillations in the context of optics, where some clear examples of the power of the superoscillatory lens (SOL) for superresolution will be provided.

\section{Weak value formalism}
\label{Weak value formalism}
Aharonov's scheme can be understood as a generalisation of the definition of a typical Von Neumann measurement in quantum mechanics. The Von Neumann Hamiltonian, $H=g(t)PC$, describes the momentum-coupling $g(t)P$ between the state and the pointer variable $Q$ \cite{aharonov2008vector, Aharonov2005paradoxes}.

\begin{figure}
    \centering
    \includegraphics[scale=0.7]{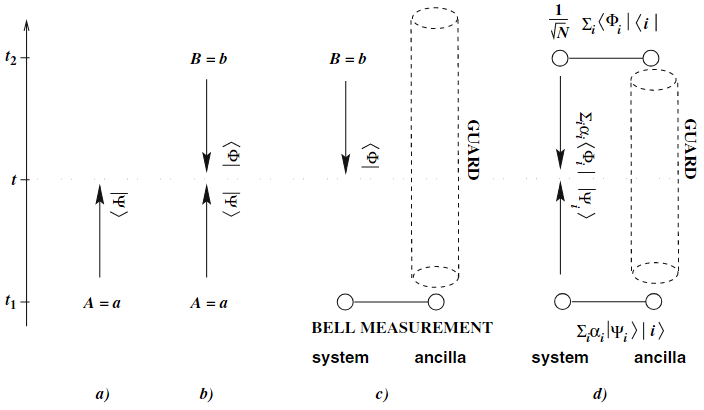}
    \caption{Description of quantum systems with the two-state vector formalism along time axis where $t_1 < t < t_2$. \textbf{a}) pre-selected state $\ket{\Psi}$ evolving ``forwards" in time , \textbf{b}) pre and post-selected states meeting at time $t$ (present) from different directions, \textbf{c}) post-selected state $\bra{\Phi}$ evolving ``backwards" in time, \textbf{d}) generalised pre- and post-selection for larger systems with complete sets of eigenstates $\sum_i\ket{\Psi_i}$ and $\sum_i\bra{\Phi_i}$ where $\{\ket{i}\}$ denotes the set of orthonormal states of the pointer and $1/\sqrt{N}$ is a normalisation constant for the measurement. Figure adapted from \cite{aharonov2008vector}.}
    \label{Two-state vector diagram}
\end{figure}

To understand the nature of this measurement, take an arbitrary Hermitian operator $\hat A$ with a complete set of eigenvectors $\ket{\phi_j}$ and eigenvalues $a_j$ at time $t$. In the standard description of quantum theory, a measurements outcome is defined as the expectation value of the observable $\hat A$ applied onto state $\Psi$:
\begin{equation}
\expval{\hat A}{\Psi} = \sum^{}_{j}a_j\,\lvert\braket{\Psi}{\phi_j}\rvert^2 
\label{expectation value}
\end{equation}
which will return operator eigenvalues $a_j$ with a given probability $\lvert\braket{\Psi}{\phi_j}\rvert^2$. The key here is such a measurement determines the state of the system in the past, relative to time $t$. This time asymmetry can be removed by implementing the two-state vector formalism introduced by Watanabe \cite{Watanabe1955asymmetry} and further developed by Aharonov et al \cite{Aharonovspin,aharonov2008vector}. Now, a system at a given time $t$ will be completely described by a two-state vector: $\bra{\Phi}\; \ket{\Psi}$
where $\ket{\Psi}$ is the state defining the measurement performed on the system in the past relative to $t$ and $\bra{\Phi}$ is a quantum state evolving backwards in time after time $t$. As such, $\ket{\Psi}$ and $\bra{\Phi}$ are commonly called the ``pre" and ``post"-selected states. This can be generalised similar to (\ref{expectation value}) now with a superposition of two-state vectors spanning the whole eigenspace. These different considerations are shown in Fig.~\ref{Two-state vector diagram}. 

The most intriguing prospect of this two-state vector formalism concerns the phenomenon known as ``weak measurement". The Hamiltonian for this measurement is equally described by the Von Neumann Hamiltonian, now with a weakened coupling\cite{aharonov2008vector, Aharonov_2011_properties, Aharonov2005paradoxes}. The idea behind this is to make the state-pointer coupling sufficiently weak so that the change in the quantum state due to its measurement can be neglected, as opposed to the Von Neumann, ``strong", measurements where the system is noticeable altered. Indeed, similar to (\ref{expectation value}), the outcome of a weak measurement of an observable $A$ given by a two-state vector $\bra{\Phi}\; \ket{\Psi}$ is given by its ``weak value":

\begin{equation}
A_w=\frac{\bra{\Phi}A\ket{\Psi}}{\bra{\Phi}\ket{\Psi}}    
\label{weak value}
\end{equation}
Unlike the expectation value, the weak value consists of real and imaginary components\cite{jozsa2007complex}. The pointer reading will return the real part while the imaginary part can be extracted by looking at the shift in momentum. However, what is particularly interesting is what happens to (\ref{weak value}) when almost orthogonal pre and post-selected states are considered. As the denominator tends to zero, the weak value can be arbitrarily large, and indeed go beyond its originally bounded eigenvalue spectrum. This is exactly the sort of behaviour that is presented in a seemingly paradoxical way in the original work from Aharonov \cite{Aharonovspin}. With reference to (\ref{weak value}), the claim of a spin-half particle's spin component being 100 no longer seems as implausible. 

Importantly, these results are not limited to the complex quantum world, as the underlying mathematics hints at an even deeper mathematical phenomenon \cite{aharonov1990superpositions}. With aid of the TSVF it is possible to construct superpositions of time evolutions (pre and post selections with certain conditions) of a quantum system during interval $T$, corresponding to different Hamiltonians with given parameters $\{a_i\}$. It can then be shown, \cite{aharonov1990superpositions}, this superposition may yield, effectively, a single time evolution corresponding to a value of the parameter $a'$ which is far outside the range of $\{a_i\}$. 

As abstract as this effect may seem, it is easier to visualise once understood its dynamic consequence. Taking the Hamiltonian to be the force acting along an axis on the system, it's wavefunction can be obtained. This wavefunction $\Psi$ will now experience a ``shift" in the direction (say $p$) of the force with normalisation constants $\{c_i\}$ and Hamiltonian parameters $\{a_i\}$

\begin{equation}
    \sum_i c_i \Psi\left(p - a_iT\right)
\end{equation}
It is then possible to find a parameter $a'$ such that 

\begin{equation}
    a'\equiv \sum_i c_ia_i \longrightarrow  \sum_i c_i \Psi\left(p - a_iT\right)\cong \Psi\left(p - a'T\right)
    \label{weak strong forces}
\end{equation}
which will only hold true if $a'$ is the weak value of the system. Equation (\ref{weak strong forces}) above makes the superposition of ``weak" forces $a_i$ acting for a short period of time equivalent to a ``strong" force $a'$ acting for the same period of time\cite{aharonov1990superpositions}. This is a form of amplification, since $a'$ is outside the spectrum $\{a_i\}$, which is peculiar to quantum mechanics. Though no classical connection could be found at first\cite{aharonov1990superpositions}, it was later realised \cite{berry1994faster} this effect could be explained via surprising interference phenomena known as superoscillations, removing the quantumness of the result. Such superoscillations are the result of band-limited functions locally oscillating faster than their fastest Fourier component\cite{berry1994faster}. The connection with the effect above now becomes quite evident as $\{a_i\}$ takes the role of bounded frequency components of the function and $a'$ is a frequency that is locally (i.e. very short time period) amplified beyond the largest value of the original spectrum.

The next section will be concerned with presenting these ``superoscillating" functions.

\section{Construction of superoscillatory functions}

A band-limited function is a function with bounded frequency components in its Fourier representation,
 \begin{equation}
     f(x)= \frac{1}{\sqrt{2\pi}}\int_{-k_\text{max}}^{k_\text{max}}\diff k\, \tilde{\!f}(k)\e^{\text{i}kx}
 \label{band-limited}\end{equation}
 its fastest allowed variation is defined by $\exp(\pm\text{i}k_{\text{max}}x)$. However, there are instances in which the function can exceed this limit, and, indeed, do so over arbitrarily large intervals \cite{Kempf2000Black_Holes, Berry_2006_time}, which will be the focus of this paper. Construction of these superoscillatory functions proves to be relatively simple, with some common examples shown below.
 \subsection{Elementary superoscillations}
 Consider the most common example of such a function\cite{Berry_2006_time, Aharonov_2011_properties}: 
 \begin{equation}
     f(x) = \left(\cos\left(\frac{x}{N}\right)+ \text{i}a\sin\left(\frac{x}{N}\right)\right)^N \quad a>1 , N \gg 1
 \label{Elementary SO}\end{equation}
 with period $\Delta x=N\pi$. For $a=1$ this returns a standard plane wave. In the limit $x\ll1$, the series expansion yields: 
 \begin{equation}
     f(x)\approx \exp\left(N\log\left(1+\text{i}a\frac{x}{N}\right)\right)\approx\exp(\text{i}ax) \quad \text{with wavenumber }a
 \label{Taylor series SO}\end{equation}
 
 However, looking at the Fourier series representation of (\ref{Elementary SO}):
 \begin{equation}
     f(x)=\sum_{m=0}^{N}c_m\exp(\text{i}k_{m}x) \,, \quad |k_m| \leq 1 
 \label{Fourier series SO}\end{equation}
 where\begin{equation*}
     c_m=\binom{N}{m}\left(\frac{1+a}{2}\right)^{N-m}\left(\frac{1-a}{2}\right)^{m}=\frac{N!}{2^N}(-1)^N\frac{(a^2-1)^{N/2}\left[(a-1)/(a+1)\right]^{Nk_m/2}}{\left[N(1+k_m)/2\right]!\left[N(1-k_m)/2\right]!}
 \end{equation*}
 and the frequency components $k_m$ are explicitly band-limited. Importantly, $c_m$ will alternate in sign, which raises the possibility of such functions being the result of near perfect destructive interference between their Fourier terms.

 Comparison between (\ref{Taylor series SO}) and (\ref{Fourier series SO}) requires all Fourier components $k_m$ to be smaller than $a$, namely: \begin{equation}
     k_m=1-\frac{2m}{N}
     \label{SO frequency}
 \end{equation}
 resulting in superoscillations with a factor $a$ faster than ``normal" oscillations, which may be chosen to be arbitrarily large. It can be seen (\ref{Elementary SO}) contained only very large wavelengths $N$, however, the expression for the superoscillatory region, $x\ll 1$, has a wavelength $1/a$ that can be much smaller. This implies localised fast oscillations will not be apparent from the global Fourier transform, but rather will only arise when considering these fast oscillatory regions.
 
 \begin{figure}[t]
     \centering
     \includegraphics[scale=0.4]{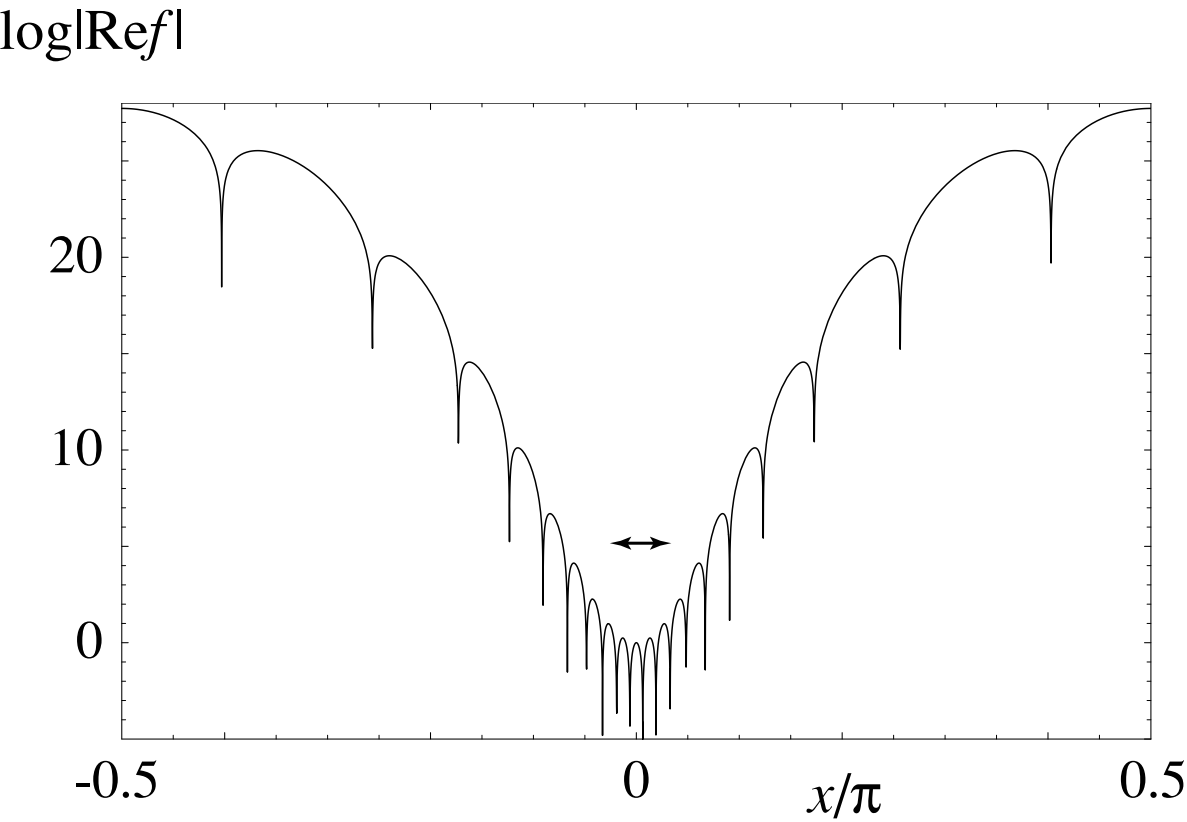}
     \caption{Superoscillatory function from (\ref{Elementary SO}) over one of its periods. The horizontal double ended arrow indicates the fastest Fourier component of the band-limited function, with sub-wavelength scale variations in the superoscillatory regime. Outside of this central region the function grows to very large values, indicative of the highly sensitive near-perfect destructive interference required to form superoscillations. Figure adapted from \cite{Berry_2006_time}}
     \label{Gaussian SO plot}
 \end{figure}
 
 \subsection{Arbitrary superoscillations by Fourier synthesis}

 The small $x$ behaviour of superoscillations like that of  (\ref{Elementary SO}) is key to reproduce small detail of arbitrary shape. To achieve, for example,  a superoscillatory narrow Gaussian of width $L$ it suffices to change the exponential present in its integral representation with any superoscillatory function with the same small $x$ behaviour\cite{berry2016synthesis}:

 \begin{equation}
   g(x)=  \frac{1}{\sqrt{2\pi}L}\exp\left(-\frac{x^2}{2L^2}\right) = \frac{1}{2\pi}\int_{-\infty}^{\infty}\diff a \exp\left(-\frac{1}{2}a^2L^2\right)\exp(\text{i}ax)
\label{Target gaussian}
 \end{equation}
 in the limit $L\rightarrow 0$, $g(x)\rightarrow\delta(x)$.
 
 Then Fourier-synthesizing $\exp(\text{i}ax)$ from (\ref{Target gaussian}) as specified, this yields a Gaussian of sub-wavelength width $L$ and arbitrarily small detail in its superoscillatory region. In \cite{berry2016synthesis} the superoscillatory function chosen is a solution to the 3D Helmholtz wave equation first presented in \cite{Berry_1994billiards}, band-limited in each of its coordinates. This then leads to:
 \begin{equation}
     g(x,\Delta)=\frac{1}{\pi\Delta}\exp\left(-\frac{1}{\Delta}\right)\int_{-\infty}^{\infty}\diff a \exp\left(-\frac{1}{2}a^2L^2\right)\frac{\sin R(x,\Delta)}{R(x,\Delta)}
\label{SO Gaussian} \end{equation}
 where $\Delta\ll1$, $a>1$ and $R$ is the radius containing the three different coordinate components of the Helmholtz solution.
 
 Although Gaussian in shape, this superoscillatory function will grow to very large values outside of its narrow superoscillatory region which will hide the Gaussian curve unless a similarly narrow range is looked at, as shown in Fig. \ref{Gaussian SO plot}. 
 
 One of the conclusions that can be derived from such a phenomenon concerns its implications for signal processing, where low pass filters (LPF) can be used to attenuate undesired frequencies higher than a cutoff frequency. Therefore, superoscillations in a function $F(t)$ emerging from a perfect low pass filter can generate the illusion the filter in question is leaky \cite{berry2019roadmap}. Precisely, what seems to be a large function with distinct zeroes and peaks once band-limited by the LPF (Fig. \ref{Gaussian SO plot}a) can show very fine detail at small scales, i.e. superoscillations (Fig. \ref{Gaussian SO plot}d). 
 
 Similarly, it is possible, in principle, to use superoscillations as a means of bandwidth compression. In the most common example mentioned in \S1, Berry postulates it should be possible to perfectly generate Beethoven's 9th symphony (one hour long, needing up to 20KHz for good HiFi reproduction) with a 1Hz band-limited signal\cite{berry1994faster}. Of course this comes at a price, as outside this superoscillatory range the function is amplified by a factor $10^{19}$ \cite{berry1994faster, Kempf2000Black_Holes}. This is because the Fourier components that are required to achieve the near perfect destructive interference associated with superoscillations are vast.
 
  \begin{figure}[t]
     \centering
     \includegraphics[scale=0.45]{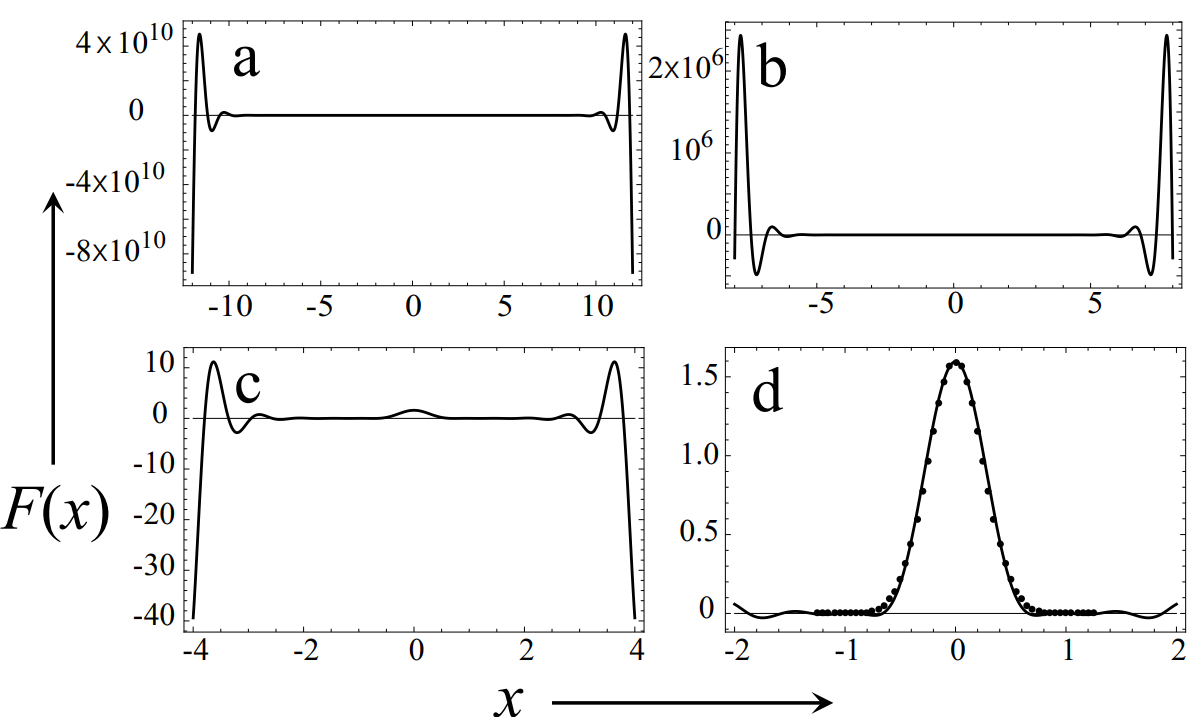}
     \caption{Computed Gaussian from (\ref{SO Gaussian}) with $\Delta=0.005$ and $L=0.25$ \textbf{a - b})Superoscillations hidden in the very broad range of the function, \textbf{c}) Gaussian starts to emerge when zooming in to a comparatively small range , \textbf{d}) Full target Gaussian (\ref{Target gaussian}), can be seen in the small $x$ regime with an amplitude 10 factors smaller than what the function grows to in the sidelobes. Adapted from \cite{berry2016synthesis}}
     \label{Gaussian SO plot}
 \end{figure}
 
 \subsection{Superoscillatory functions by interpolation}
Building from Berry's Beethoven signal compression example above, Kempf and Ferreira show a more general and surprising result\cite{Kempf2006Nyquist}: given a finite-energy signal band-limited to any arbitrarily small frequency, say 1Hz, it is possible for it to be interpolated through specified points on any waveform of arbitrarily large frequency. Therefore, this allows the construction of a superoscillatory function whose shape can be controlled. The interpolation function can contain arbitrarily fast oscillations by specifying arbitrarily close points. More formally\cite{Kempf2000Black_Holes}, given arbitrarily chosen times $\{ t_i\}_{i=1}^{N}$ and amplitudes $\{ A_i\}_{i=1}^{N}$, with $N\gg1$, a function can be found satisfying \begin{equation}
    f(t_i)=A_i
\label{SO interpolated}\end{equation}

Its energy is defined as the norm-square of the amplitude, and so the energy ``cost" of such superoscillations can be estimated. It is established \cite{Kempf2006Nyquist} the total energy of a superoscillatory function grows exponentially with the number of superoscillations, and polynomially with the speed (the reciprocal of the period) of the superoscillation. Exponential growth would greatly impact the practicality of superoscillations for optical applications; the polynomial relation, however, is a more manageable limitation. Further, it is possible to define a unique minimum-energy function which interpolates all the desired points\cite{Kempf2006Nyquist}. For completeness, this function $f$, band-limited to $\mu/2$ Hz, is given by:
\begin{equation}
    f(t)=\mu\sum_{k=1}^{N}a_k\sinc[\mu(t-t_k)]
\end{equation}
This approach is largely based on Shannon's original proof concerning the best approach to sampling in signal processing\cite{marks2012SamplingIntro}. It sets a sufficient condition on the sampling rate (Nyquist-Shannon limit) to recreate a continuous signal from discrete interpolation points and where aliasing effects are not noticeable (distinct signals remain distinguishable).

Total energy, however, is not the best way to represent the energy available in the usable part of the superoscillatory regime due to the inherent sidelobes of these functions. As such, energy ``yield" of superoscillations, $Y$, can be defined as the proportion of energy in the superoscillatory region of a function, $f(t)$, with respect to the total energy \cite{Katzav2013Yield}:
\begin{equation}
    Y=\frac{\int_{-a}^{a}|f(t)|^2\, \diff t}{\int^{-\pi}_{\pi}|f(t)|^2\, \diff t}
\label{Energy yield}\end{equation}
where $[-a,a]$ is the superoscillatory region of the function bounded by one full period length 2$\pi$. By specifying $M$ interpolation points then a superoscillatory function with $M-1$ arbitrarily fast oscillations can be controlled as the one in (\ref{SO interpolated}), only now the function is specified to be periodic. Applying such constraints to a periodic function, a truncated cosine Fourier series with $N+1$ terms is taken in \cite{Katzav2013Yield}, the problem reduces to an eigenvalue problem with $M$ linear equations and $N+1$ unknowns, solved numerically to find the optimal energy ratio. Maximising (\ref{Energy yield}) then becomes of great interest and indeed, optimising superoscillations is challenging due to the precision needed, but their storage for use is not as sensitive, which is important if they are to be applied onto a system where noise and error are inevitable, like a superoscillatory lens \cite{Katzav_2016statistics}. It has been found superoscillations become highly unstable if this noise is represented as a disturbance in the phase of the wave \cite{Berry_2016noise}, further establishing that, on paper, superoscillatory behaviour is very delicate. Noise in actual wave systems however, is more complex and can be studied to understand how it may show itself in the superoscillatory signal. Coherent noise, for example, has been well documented and methods like phase-sensitive optical time domain reflectometry ($\phi$OTDR) exist to suppress effects as small as a 0.4\% amplitude disturbance\cite{Martins2013Noise}. 
\label{superoscillations and noise}
 \subsubsection{Superoscillatory wavefunctions}

\begin{figure}
    \centering
    \includegraphics[scale=0.55]{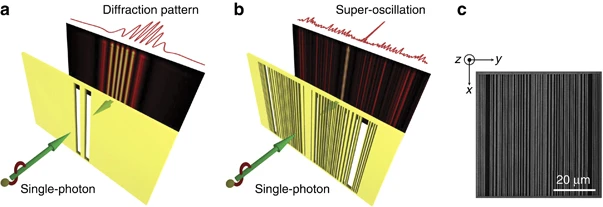}
\caption{Single photon interference from slit. \textbf{a}) Classical observation of the diffraction pattern for a single photon when passed through Young's double slit. \textbf{b}) Quantum superoscillations arising from the photon's wavefunction concentrating in the superoscillatory regime due to the slit-mask. \textbf{c}) Scanning electron microscope image of the optimised superoscillatory mask. Adapted from \cite{yuan2016photon}}
    \label{single photon SOL}
\end{figure}

This approach can also be applied to quantum mechanical effects, where superoscillations will be observed and represented by physical phenomena. For a particle wavefunction $\psi$ explicitly band-limited in momentum\cite{kempf2004unusual}, its Fourier representation in position space is \begin{equation}
     \psi(x)=\frac{1}{\sqrt{2\pi\hbar}}\int_{-p_\text{max}}^{p_\text{max}}\diff p\,\widetilde{\psi}(p)\exp\left(\frac{\text{i}xp}{\hbar}\right)
     \label{SO wavefunction}
 \end{equation} 
which is in the standard form of (\ref{band-limited}) with a shortest wavelength $\lambda_{\text{min}}=h/p_\text{max}$.%

Therefore, (\ref{SO wavefunction}) is, effectively, a finite-energy (it is normalised) band-limited signal (in momentum). This allows, again, to construct a superoscillatory function by specifying points for the wavefunction to pass through by the same method that resulted in (\ref{SO interpolated}). In which case, the wavefunction will contain sub-wavelength oscillations in a narrow region of very small amplitude  compared to the large sidebands outside of it. The probability of finding the wavefunction at a certain position is given by its modulus squared. It follows the ``cost" of having superoscillations in the wavefunction is that it suppresses the probability of finding the particle in that region, as the contribution from it will be minimal in comparison. As always, it is in these regions where highly sensitive phenomena will become apparent. To further motivate this a physical example can be considered where the particle's properties in this region will be explored, e.g. its behaviour when passed through a slit\cite{kempf2004unusual, kempf2005wavefunctions}. 

Consider a particle incident onto a screen with a single slit in such a way that it is only the superoscillatory part of the wave function which passes through the slit. Immediately behind the other side of the slit will then emerge a wavefunction with very rapid oscillations, that is zero elsewhere. These sub-wavelength oscillations now do show as high frequencies in its Fourier representation, as opposed to (\ref{band-limited}). This is because the delicate equilibrium once present between the contributions from the fast oscillations in the slit interval and the contributions from outside it cancelling, no longer exists, since the latter is taken to be zero. Therefore, the emerging quantum particle obtains a correspondingly high momentum expectation value \cite{kempf2004unusual}. This seems to indicate the particle gains momentum by just passing through the slit. The gain in momentum will be related to the superoscillatory wavefunction's wavelength and, as before in (\ref{SO frequency}), this can be chosen to be arbitrarily large. 

These large momentum impulses therefore could be measured through Aharonov's quantum weak measurement scheme \cite{Aharonovspin ,Aharonov2005paradoxes, aharonov2008vector} where the  weak value of the wavenumber (expectation value of momentum operator~\cite{berry2013momenta}) is determined. A demonstration of this effect can be realised with a single photon \cite{yuan2016photon}, whose wavefunction is tightly localised around an suitably optimised slit mask, Fig.~\ref{single photon SOL}. Behind the mask a superoscillatory pattern will emerge, with potential applications, as will be discussed later, for far-field, focusing techniques at sub-wavelength scales. The result is, however, quite unintuitive for small ensembles as it seems to indicate, in the classical sense, the individual photon interferes with itself along its trajectory. Such an outcome would be hard to explain when performing a ``strong" measurement but, under the weak measurement scheme, the position of the photon will only be determined by a reconstruction of the photon momentum after a post-selection with no direct position measurement. Though hard to visualise, single photon interference phenomena can and do indeed happen \cite{Kocsis2011singlephoton}.

\subsection{Naturally arising superoscillations}

These last few examples have considered cases where superoscillations can be created purely from mathematical manipulation of their well studied features. Surprisingly, however, superoscillations will also arise naturally, and are, in fact, unexpectedly common. Consider a physical example: monochromatic, complex random waves in $D$ dimensions \cite{berry2008natural}. These type of waves can be understood as the superpositions of plane waves with random complex amplitudes and directions with greatest wavenumber component $k_0$, similar to (\ref{Fourier series SO}).
\begin{equation}
    \psi(\bm{r})=\sum^{N}_{n=1}a_n\exp\left(\text{i}\bm{k}_n\cdot\bm{r}\right) \quad N\gg1,\; a_n\in\mathbb{C}, \quad |\bm{k}_n|=k_0
\label{monochromatic SO function}\end{equation}
The physical significance of the wave above is that, as in $\S$3.2, they must also obey Helmholtz's equation:

\begin{equation}
\nabla^2\psi+k_0^2\psi=0
\label{Helmholtz}    
\end{equation}
Recall from (\ref{phase gradient}) the phase gradient is a natural measure of the oscillation at a given point $\bm r$; the wavenumber. Superoscillations correspond to regions of the wave where it is varying at sub-wavelength ($2\pi/k_0$) scales. This will happen at the optical vortices of the wave, or phase singularities, where the phase gradient of (\ref{monochromatic SO function}) diverges to arbitrarily large values. 

The phase gradient, $\nabla\arg\psi(\bm{r})$, is also related to another common term in wave physics, more intuitive in this context, the current density \cite{dennis2008superoscillation}: 
\begin{equation}
\bm{J}(\bm{r})=\text{Im} [  \psi^*(\bm{r})\nabla\psi(\bm{r})]
\label{current density}
\end{equation} 
which itself leads to an equivalent construction for the momentum (wavenumber) of the wave \cite{berry2013momenta}, that can be similarly used to measure superoscillations:

\begin{align}
\begin{split}
    k_\text{current}(\bm{r})=\frac{\bm{J(r)}}{|\psi(\bm{r})|^2}=\frac{\text{Im}[  \psi^*(\bm{r})\nabla\psi(\bm{r})]}{|\psi(\bm{r})|^2}= \text{Im}\frac{\nabla(\psi(\bm{r})}{\psi(\bm{r})}=\nabla\text{Im}[\log\psi(\bm{r})] & = \nabla\arg\psi(\bm{r}) \\
   & = k_\text{phase}(\bm{r})
   \end{split}
\end{align}

Now, the simplest measure of superoscillations in the wave is the probability, $P_{D_{\text{super}}}$, that $k(\bm{r})>k_0$ at an arbitrary point $\bm{r}=\{x_1,x_2..., x_D\}$. That is:
\begin{equation}
    P_{D_{\text{super}}}= \int_{k_0}^{\infty} \diff{k}P_D(k)
\label{SO probability integral}
\end{equation}

\begin{figure}[t]
    \centering
    \includegraphics[scale=0.5]{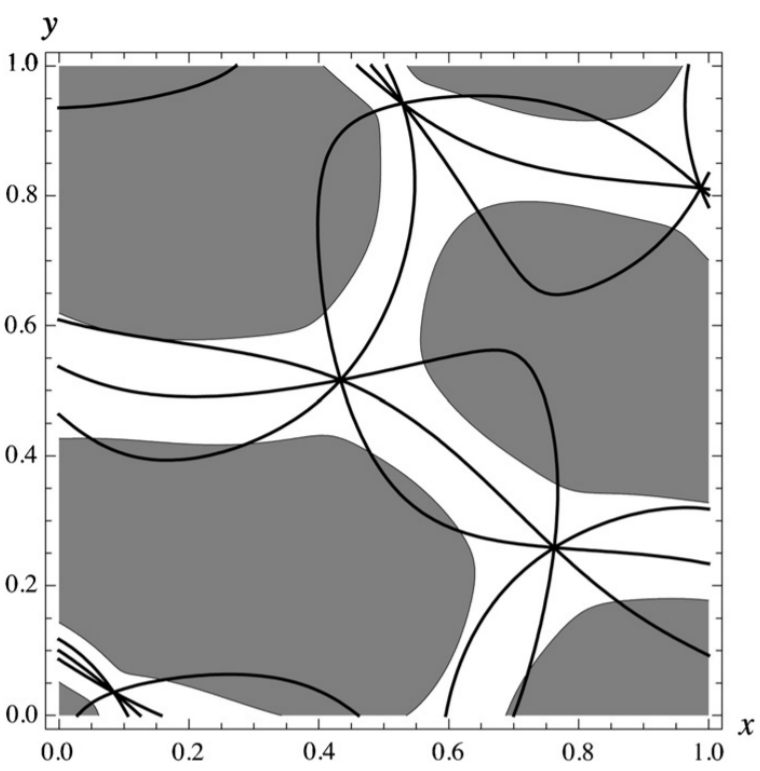}
    \caption{Superoscillations in D=2 arising from plane wave superposition from (\ref{monochromatic SO function}) with $N=10$, $k_0=2\pi$. Shows wave vortices (phase singularities) where equally separated wavefronts intersect (black lines) with superoscillatory regions ($k(\bm{r})>k_0$) in white. Surprisingly, large fractions of the wave are naturally superoscillatory, with no need of artificial construction. Figure adapted from \cite{berry2008natural}}
    \label{monochromatic wave SO plot}
\end{figure}

In \cite{dennis2008superoscillation} it is shown in the case of $D=2$, $1/3$ of the area fraction of a speckle pattern will be superoscillatory. The generalisation of this result is given in \cite{berry2008natural}: 
\begin{equation}
        P_{D_{\text{super}}}= 1-\left(\frac{D}{D+1}\right)^{D/2}
\label{SO probability}
\end{equation}
With values ranging from  $P_{1_{\text{super}}}\simeq0.29289$ for $D=1$ to the case where $D\rightarrow\infty$ with $P_{\infty_{\text{super}}}\simeq0.39347$. Showing that superoscillations are a common feature of any wave, making it possible for them to be used for practical applications. 

A similar, even more general, calculation is reproduced in \cite{berry2010superweak, berry2011spin} where, instead of looking at regions of fast oscillations in waves, the shift of a pointer state beyond its allowed region due to weak value amplification is considered. This scheme enables discrete superoscillatory wavenumbers beyond its spectrum (weak values) to be found. The connection between these two phenomena will become more apparent after the next section.
\label{discrete SO probability}

\section{Quantum mechanical framework of superoscillations}
\subsection{Pointer shifts in a weak measurement}
It was shown in (\ref{weak value}) that, by weakening the ordinary von Neumann interaction Hamiltonian, the weak value of a two-state vector follows, with some peculiar properties. Mainly, the weak value can lie outside the expected spectrum of its observable. Despite the term weak, these measurements can be quite precise. In \cite{Aharonovspin} this property is explored in reference to a large number $N$ of spin-1/2 particles' pre and post-selected spin states, $\ket{\hat{S}_z=\frac{N}{2}}=\prod_{j=1}^{N}\ket{\uparrow_z}_j$ and $\ket{\hat{S}_x=\frac{N}{2}}=\prod_{j=1}^{N}\ket{\uparrow_x}_j$, respectively.

In the weak regime a measurement can then be made up to an uncertainty of $\sqrt{N}$ without disturbing the system itself by more than $\sqrt{N}$. One way to do so might involve measuring the spin along a direction $\xi=45^\circ$ relative to the $x$-$z$ plane \cite{Aharonov_2011_properties}, modelled by the collective observable $\ket{\hat{S}_\xi^{(N)}}\equiv\frac{1}{N}=\sum_{i=1}^{N}\left(\frac{\hat{S}_x^i+\hat{S}_z^i}{\sqrt{2}}\right)$ which can then be inserted into (\ref{weak value}):

\begin{equation}
    \hat{S}_{\xi,\;w}^{(N)}=\frac{\prod_{k=1}^{N}\bra{\uparrow_z}_k\left(\hat{S}_z^{(N)}+\hat{S}_x^{(N)}\right)\prod_{j=1}^{N}\ket{\uparrow_x}_j}{\sqrt{2}\left(\bra{\uparrow_z}\ket{\uparrow_x}\right)^N} = \frac{N/2+N/2}{\sqrt{2}}=\frac{\sqrt{2}}{2}N\pm O\left(\sqrt{N}\right)   
\label{weak value spin}
\end{equation}
where the second step is allowed since both $\hat{S}_z^{(N)}$ and $\hat{S}_x^{(N)}$ can be independently measured in this setup without disturbance to the system, they commute.
While the possible values for $\hat{S}_{\xi}^{(N)}$ in a spin-1/2 particle are bounded by its eigenvalues $\pm \frac{N}{2}$, (\ref{weak value spin}) returns values completely outside this spectrum. 

This offers an explanation to the system's state in a weak measurement. Superoscillations however, are an effect that can be perceived by the measuring device or pointer rather than the quantum system itself. The reason as to why the pointer will shift beyond its bounded spectrum into, apparently, forbidden values has to do with, as mentioned before, the Fourier components of the pointer state interfering constructively around these, amplifying its range. 

So, generalising what was outlined above, for an arbitrary bounded operator $\hat{A}$, an initial pointer state $\ket{\phi}$ can be introduced which, along with the pre-selected state $\ket{i}$ defines the pointer-system combined initial state $\ket{i}\ket{\phi}$. The measurement will be briefly coupled to the momentum $\hat{p}$ of the pointer, with strength $\lambda$ \cite{berry2011pointer}. Projecting now the whole system onto a post-selection $\ket{f}$, the final pointer state is given by:

\begin{equation}
    \ket{\psi}=\bra{f}\underbrace{\exp(-\text{i}\lambda\hat{A}\hat{p})}_\text{pointer shift}\ket{i}\ket{\phi}
    \label{pointer state}
\end{equation}

where $\hat{A}$'s eigenvalues, $A_n$, are bounded and given by:

\begin{equation}
    \hat{A}\ket{n} = A_n\ket{n}, \quad A_\text{min}\leq A_n \leq A_\text{max}
\end{equation}

This post-selection can be understood as the quantum mechanical analogue of what statisticians call Large Deviation Theory. Formalised by Varadhan \cite{varadhan1966asymptotic}, the concept behind it is to look at a very small fraction of the results of an experiment corresponding to very rare events at the remote tails of the probability distribution, hence the name weak. 

In position, $q$, representation: $\hat{p}=-\text{i}\hbar\partial_q; \; \ket{\phi}\rightarrow\bra{q}\ket{\phi}=\phi(q),\; \ket{\psi}\rightarrow\bra{q}\ket{\psi}=\psi(q)$. So (\ref{pointer state}) becomes:

\begin{align}
    \psi(q) &= \bra{f}\exp(-\text{i}\lambda\hat{A}(-\text{i}\hbar\partial_q))\ket{i}\phi(q)\notag\\
    &= \sum_{n=-N}^N\underbrace{\bra{f}\ket{n}\bra{n}\ket{i}}_{\text{a constant}}\phi(q-\lambda\hbar A_n)\notag \\
    &= \sum_{n=-N}^N C_n\phi(q-\lambda\hbar A_n)
    \label{shifted pointer}
\end{align}
where $C_n$ is a measure of the overlap between the pre and post-selection and the eigenstates of the pointer.

An important result follows from (\ref{shifted pointer}): the final state of the pointer, unsurprisingly, is a superposition of copies of its initial state, shifted by eigenvalues $A_n$. This comes, however, with a crucial observation. Commonly in quantum mechanics, $\ket{f}=\ket{i} \rightarrow C_n=\lvert \bra{n}\ket{i}\rvert^2$ (no two-state vector). Nevertheless, if the pre and post-selected states are different, $C_n$ can have different phases and so the overlaps can interfere destructively. In turn this generates a pointer shift large enough so as to push the final pointer state beyond its allowed range, albeit greatly reduced in strength (weak)\cite{berry2011pointer}. More precisely, if the Fourier transform $\tilde{\phi}(q)$ decays fast enough, i.e. if the pointer state is broad enough so that it covers all the eigenstates, then the pointer wavefunction, almost cancelled by near-perfect destructive interference, is resurrected far from the spectrum. This ``supershift" is characterised by its weak value:
\begin{equation}
    \psi(q)\approx\bra{f}\ket{i}\phi(q-\lambda\hbar\text{Re}A_w)
\end{equation}
which will be arbitrarily large when the pre and post-selected states in (\ref{weak value}) are nearly orthogonal. The imaginary part of the weak value is not of great relevance for this particular discussion but it similarly has some interesting properties\cite{jozsa2007complex}. Namely, this setup is partly simplified by taking the initial and final pointers to be static, in which case $\text{Im}A_w=0$. However should the pointer be moving, then $\text{Im}A_w\neq0$ and its movement would be characterised by the initial pointer momentum variance:
\begin{equation}
    \expval{p}=2\lambda\text{Im}A_w\expval{\hat{p}^2}{\phi}
\end{equation}
which will be quite small if the pointer is broad, as this scheme requires it to be.

\begin{figure}[!t]
    \centering
    \includegraphics[width=110mm, height=100mm]{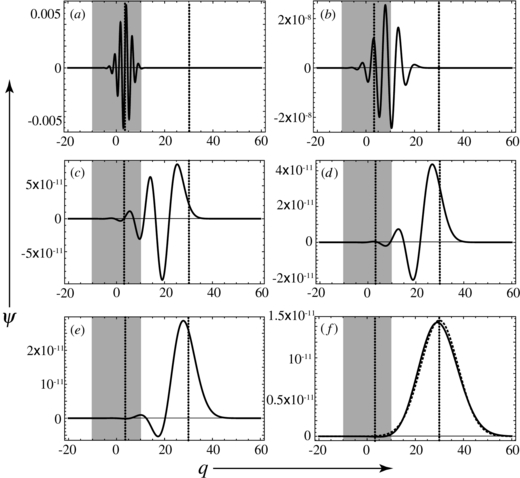}
    \caption{Pointer wavefunctions as a function of position, $q,$ after a weak measurement for initial Gaussian pointer states as in (\ref{Gaussian pointer}) of widths, $\Delta$, regularly increasing  from \textbf{a})$\Delta$=1 to \textbf{f})$\Delta$=10. The spectral range of the initial state is $|N|\leq10$, the shaded region. A narrow pointer, \textbf{a}), will have oscillations entirely localised within $N$. As the width of the pointer is increased, \textbf{b})-\textbf{e}) the wavefunction gradually leaks out from $N$, until the pointer wavefunction is fully located outside the spectrum, \textbf{f}). Note the Gaussian is normalised which results in much weaker values once the pointer is supershifted, with the peak in \textbf{f}) 8 orders of magnitude smaller than its counterpart in \textbf{a}). Figure adapted from \cite{berry2010superweak}}
    \label{pointer supershift plot}
\end{figure}

An illustration of (\ref{shifted pointer}) is shown in Fig.~\ref{pointer supershift plot} for an initial pointer state with Gaussian shape:
\begin{equation}
    \phi(q)=\frac{1}{\Delta\sqrt{2\pi}}\exp{-\frac{q^2}{2\Delta^2}} \longrightarrow \quad \tilde{\phi}(p)=\frac{1}{2\pi}\exp{-\frac{\Delta^2p^2}{2}}
    \label{Gaussian pointer}
\end{equation}

\subsubsection{Statistics of pointer supershifts}
This effect's strength is analysed in \cite{berry2010superweak} where the probability distribution of the weak values is considered, building from previous work in $\S$\ref{discrete SO probability}. For an operator, $\hat{A}$, with eigenvalues bounded by $-A_\text{max}\leq A_n \leq A_\text{max}$
\begin{equation}
    P_\text{super}=\int_{-\infty}^{-A_\text{max}}\diff{A}P(A)+\int_{A_\text{max}}^{\infty}\diff{A}P(A)
    \label{superweak probability}
\end{equation}

It is found the weak value probability is simply a Cauchy distribution, with simplifies (\ref{superweak probability}) to:

\begin{equation}
    P_\text{super}=2\int_{A_\text{max}}^{\infty}\diff{A}P(A) = 1 - \frac{A_\text{max}}{\sqrt{\expval{A_n^2}+A^2_\text{max}}}
\end{equation}
which then depends only on the eigenvalue distribution for $A$. When $\expval{A_n}$ is concentrated at its eigenvalues $\pm A_\text{max}$ the probability of a pointer supershift is $P_\text{super}=1-1/\sqrt{2} \simeq0.29289$. This is the same result that was found for superoscillations in the case of 1D monochromatic waves in (\ref{SO probability}), making the link between the two even more apparent. Further, if $\expval{A}$ has only two discrete eigenvalues, the probability distribution has two spikes at $\pm A_\text{max}$ and is zero elsewhere, as is the case of a spin-1/2 particle. In such a case, the probability for the weak value of the spin to lie outside of the spectrum is $P(|S|>1/2\hbar)=1/3$ \cite{berry2011spin}, and the probability for $P(|S|>100\hbar)$, to answer the problem posed by Aharonov Aharonov et al\cite{Aharonovspin}, is $\frac{1}{120000}$.

\subsection{Prediction of quantum effects near vortices}

As briefly mentioned in $\S\ref{discrete SO probability}$, optical vortices, the zeroes of the wavefields, are the clearest examples of naturally arising superoscillations and so have some peculiar properties which will now be explored with a few examples. Indeed, predictions can be made about physical effects near vortices, as the local wavevector $k\rightarrow\infty$. To aid with the visualisation it is helpful to consider the most obvious interpretation of $k$ which, once multiplied by $\hbar$, is of a local momentum. However, momentum is a dynamic property and so to see its effect an interaction is required. 

To this effect, a weak measurement on the superoscillatory $k(\bm{r}) : |k(\bm{r})| > k_0$ near a vortex corrresponds to the momentum transferred in individual photon impacts on a small detector (e.g. an atom) \cite{barnett2013superweak}. The reasoning behind this, seen in a similar context in $\S3.3.1$, is that the atom does not feel the overall Fourier transform of the waves (sum of plane waves with momentum $<k_0$), but rather it ``sees" the field only where the atom is, near the vortex.

\subsubsection{Momentum transfer near optical vortices}

A simple model for this explanation \cite{barnett2013superweak} can be constructed inserting a two-level atom as the probe into an optical field $E(\bm{r})$ with ground and excited states $\ket{g}, \ket{e}$, respectively, coupled by electric dipole interaction:
\begin{equation}
    \hat{\mu}=\mu\left(\ket{g}\bra{e}+\ket{e}\bra{g}\right)
\end{equation}
Now, making some simplifications in the model, namely assuming the atom to be effectively static and its energy levels to be in resonance, the rotating wave approximation can be used, reducing the Hamiltonian to:
\begin{equation}
\hat{H}=-\mu\left(\ket{g}\bra{e}E^*(\bm{r})+\ket{e}\bra{g}E(\bm{r})\right)
    \label{RW hamiltonian}
\end{equation}
Taking the initial position wavefunction for the atom as a Gaussian centred on $x_0$, close to the vortex cross-section $\sigma$:
\begin{equation}
    \psi_\text{init}(\bm r) = N \exp{-\frac{(x-x_0)^2+y^2+z^2}{2\sigma^2}}
\end{equation}
And the atom's initial state is given by $    \bra{\bm r}\ket{\Psi(0)} = \ket{g}\psi_\text{init}(\bm r)$,
with first order perturbation theory giving the state at time $t$:
\begin{equation}
\bra{\bm r}\ket{\Psi(t)}\approx \left(1-\text{i}\frac{\hat{H}t}{\hbar}\right)\bra{\bm r}\ket{\Psi(0)}
\end{equation}
After the transition, the motional state of the atom is $\psi(\bm r) \propto E(\bm r) \psi_\text{init}(\bm r)$. For a vortex of order $m$ optical field 
\begin{equation}
    E(\bm r) = N(x+\text{i}y)^m\exp(\text{i}k_0z)
\end{equation}
with local momentum diverging as a function of position relative to the vortex, $k(\bm r) \equiv \nabla \arg E \propto \frac{m}{r}\bm e_\phi$, the momentum representation of the motional state of the atom after the transition results in:

\begin{figure}[!t]
    \centering
    \includegraphics[width=90mm, height=80mm]{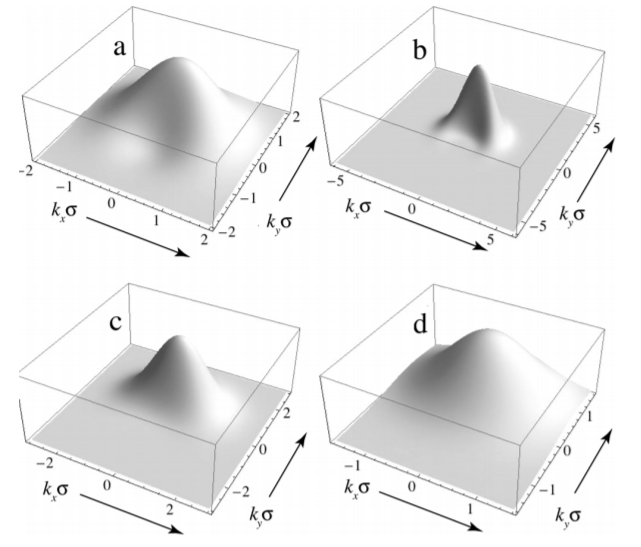}
    \caption{Momentum distribution from (\ref{momentum distribution}) of excited atom in the transverse $k_x$ and $k_y$ planes for \textbf{a}) $x_0=\sigma, m=1$, \textbf{b}) $x_0=\sigma, m=3$, \textbf{c}) $x_0=2\sigma, m=3$, \textbf{d}) $x_0=5\sigma, m=4$ showing the peak in $k_y$ shifted from $0$ further into the $k_y$ plane due to the momentum superkick. Figure adapted from \cite{barnett2013superweak}}
    \label{fig:my_label}
\end{figure}

\begin{equation}
    \tilde{\psi}(\bm k) = N \left(\partial_{k_x}+\text{i}\partial_{k_y}\right)^m\tilde{\psi}_\text{init}(k_x, k_y, k_z-k_0)
\end{equation}
Which is a redistribution of momenta the atom already possesses. A direct consequence of this is the recoil in the $z$-direction associated with the conservation of linear momentum on impact. Further, looking at the final momentum probability distribution a shift in the $k_y$ direction appears:
\begin{equation}
    \lvert \psi(\bm k)\rvert^2 = N^2\exp\left(-(k_z-k_0)^2\sigma^2\right)\left( k_x^2+ (k_y+ x_0/\sigma^2)^2\right)^m\exp(-(k_x^2+k_y^2)\sigma^2)
    \label{momentum distribution}
\end{equation}

If the atom is well localised away from the vortex, this ``superkick" reduces to the mean momentum in the $\bm {e}_y$-direction \cite{barnett2013superweak}. Again, these superkicks are rare since the wavefunction is very small near the vortices, but transverse momentum shifts are key in explaining extraordinary manifestations of optical forces at play in similar contexts \cite{picardi2018angular}.

\section{Superoscillations and superresolution}

Though stemming from purely theoretical ideas, the comprehensive concept behind superoscillations (very rapid oscillations in band-limited functions) makes for a vast range of possibilities when it comes to its applications. Though systems of interfering waves, particularly band-limited ones, are ubiquitous in many areas of science and technology, perhaps their most intriguing use concerns that of sub-wavelength imaging: superresolution.

The first attempt to overcome the classical Abbe-Rayleigh diffraction limit has often been attributed to di Francia \cite{Toraldo1952}, inspired by previous work on radar super-directivity (or super-gain) in antenna arrays. It was known at the time that if the array contains a sufficiently large number of sources their strengths and phases can be arranged such that the resulting radiation pattern is confined to an arbitrarily narrow region near the forward direction, even at the limit where the length of the array is arbitrarily smaller than the wavelength of the radiation. He devised this theory could be applied to optics where, instead of looking at radiation in the far-field, the radiation could be concentrated onto a focal spot of sub-wavelength size. The resulting amplitude pattern from these radiation sources can easily be modelled as a band-limited function, converging the studies in super-directivity with superoscillations\cite{Berry2014Endfire}. In fact, it has also been proven that any small field feature can be represented as a series of band-limited functions given a prescribed field of view. Moreover, despite their highly sensitive nature\cite{Berry_2016noise}, if unperturbed, superoscillations do not dissipate under time evolution, unlike evanescent waves, motivating the idea that superoscillations propagate sub-wavelength structure further into the field than evanescent waves\cite{Berry_2006_time}. Evanescent waves had been known to contain the fine-scale detail of the radiating object (as opposed to the propagating waves that could be classically observed), but their imaging power was inherently limited by their almost null propagation, requiring near-field optics with undesirable effects in many contexts due to its invasive nature \cite{barwick2009photon, rogers2013optical}.

\begin{figure}[t]
    \centering
    \includegraphics[scale=0.8]{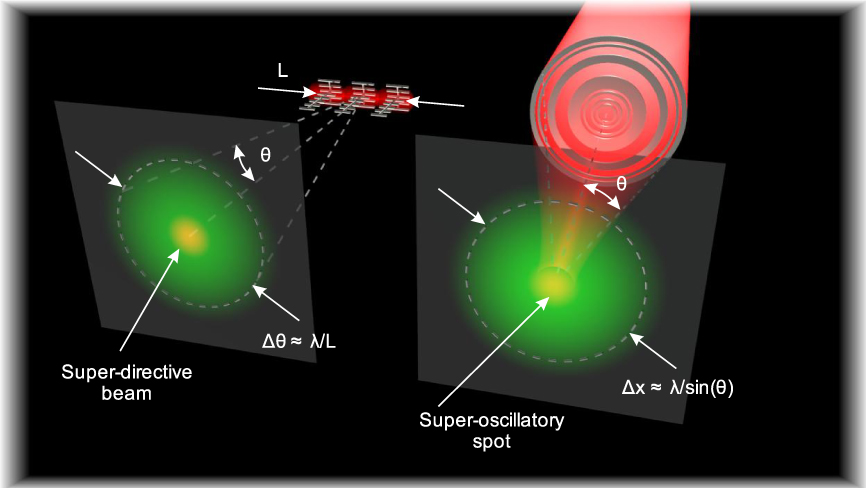}
    \caption{Interplay between antenna super-directivity and superoscillations. On the left, a super-directive antenna array generates a narrow super-directive beam beyond the diffraction limit. On the right a super-oscillatory lens generates a sub-wavelength hot-spot. Adapted from \cite{rogers2013optical}}
    \label{super-directivity}
\end{figure}

Therefore, to achieve superresolution via superoscillations a series of computationally optimised concentric rings in a standard Fresnel zone-plate can be constructed, which would result in superoscillations in an extremely narrow focal spot where sub-wavelength detail would be seen. As such, the usability of superoscillations in imaging heavily relies on the costly lens-making technology required.

\subsection{Sub-wavelength focus}
\begin{figure}
\centering
\begin{subfigure}{0.9\textwidth}
    \centering
    \includegraphics[scale=1]{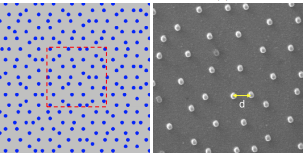}
    \caption{}
    \label{quasi-crystal}
\end{subfigure}
\begin{subfigure}{0.9\textwidth}
    \centering
    \includegraphics[scale=1]{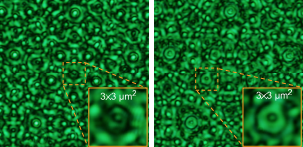}
    \caption{}
    \label{nanohole focus}
\end{subfigure}
\caption{\textbf{a}) left: Quasi-periodic sample highlighting the long range created from random pattern of holes on a plate. \textbf{a}) right: SEM image of the fragment in a quasi-periodic nanohole array. 
Bottom: Monochromatic ``Talbot carpet" arising from diffraction off a periodic grating when it is illuminated with coherent light ($\lambda$ = 500 nm) shown at different heights for image size $20\times20\mu$m$^2$. sub-wavelength spots are visible and their main features can be resolved, focal spot in the middle surrounded by bright rings. \textbf{b}) left: 7$\mu$m above array. \textbf{b}) right: 8.5$\mu$m above array. Adapted from \cite{huang2007superresolution}}
\end{figure}
Until then, similar studies explored the idea of using superoscillations to produce arbitrarily small spots on a screen. Creation of these very small spots could be created naturally through the Talbot effect\cite{Berry_2006_time, rogers2013optical}. When light is passed through a linear periodic diffraction grating, it produces a periodic and intricate pattern where many wave vortices can be observed at multiples of a constant (Talbot length). As such, very rapid oscillations in the field can be found near these points. Yet, to view superoscillatory spots (very small but not point-like), a 2D grating, or mask, is needed. To do so, a quasi-crystalline structure can be used which possesses long range order as required, but is not periodic, lending itself to a fuller study of the effect\cite{huang2007nanohole}. The choice of quasi-periodic masks should not be surprising as, in spite of the simplicity of their design, their Fourier spectrum is continuous and therefore diffraction of light from such masks generates very complex field patterns where one can expect to observe super-oscillatory features\cite{rogers2013optical}.

This quasi-crystal can be created by drilling an array of very small holes onto a plate arranged in a quasi-periodic pattern with long range structure (Fig. \ref{quasi-crystal}) from which light will diffract in a way similar to a many-hole Young's interferometer\cite{huang2007superresolution}. The resulting pattern then shows up, as expected, with regions of superoscillatory behaviour concentrated in very small spots, Fig. \ref{nanohole focus}, where the amplitude must be minute in comparison to the rest of the field. Therefore this interference approach \cite{huang2007nanohole, huang2007superresolution} was a demonstration that it is experimentally possible to construct superoscillatory spots. 

The mechanism of superoscillatory focus has been further understood more recently in terms of the energy backflow around nanoscale vortices\cite{berry2019roadmap}. Close to these, the backflow depletes the area where flow can propagate in the forward direction narrowing the focal point further beyond the conventional diffraction limit.

\subsection{Design of an optical superoscillatory lens}
However, a 2D diffraction pattern is of little use in terms of imaging. For sub-wavelength-imaging a single focal spot where all light is concentrated and superoscillates is needed, and for this an optical element must be added to the system. It is clear therefore, once established the optical capabilities of superoscillations, the main barrier to go beyond the resolution limit is the construction of a viable superoscillatory lens (SOL).
\subsubsection{Binary mask SOL}

\begin{figure}[!t]
    \centering
    \includegraphics[scale=0.4]{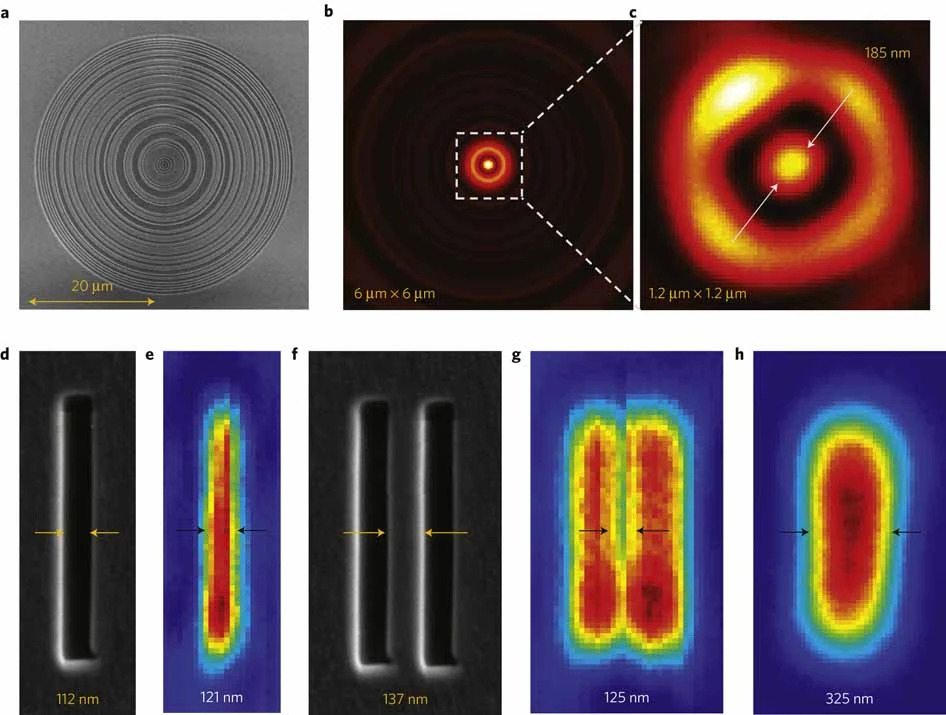}
    \caption{Sub-wavelength imaging with a SOL. a) SEM image of the ring-optimised SOL. b) energy distribution of the super-oscillatory lens at 10.3$\mu$m from the lens and c) the observed focal point when iluminated with light ($640$nm). All the main features can be seen with the intense sidelobes surrounding the sub-wavelength hotspot.
    d) SEM image of the sub-wavelength slit and e) its SOL image
    f) SEM image of the double slit separated by $137$nm and f) its SOL image with both slits clearly resolved. h) Individual slits can not be resolved with a conventional lens numerical aperture NA=1.4. 
    Adapted from \cite{zheludev2012SOL}}
    \label{binary mask SOL}
\end{figure}

\begin{figure}
\centering
\begin{subfigure}{0.30\textwidth}
    \centering
    \includegraphics[scale=0.45]{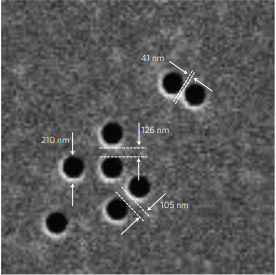}
    \caption{}
\end{subfigure}
\begin{subfigure}{0.30\textwidth}
    \centering
    \includegraphics[scale=0.45]{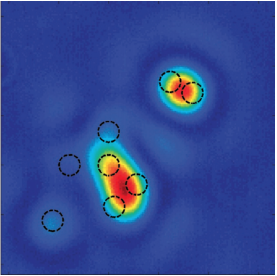}
    \caption{}
\end{subfigure}
\begin{subfigure}{0.30\textwidth}
    \centering
    \includegraphics[scale=0.45]{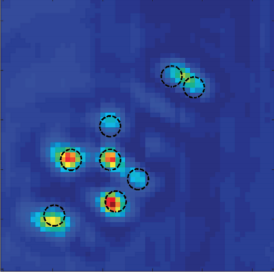}
    \caption{}
\end{subfigure}
\caption{a) SEM image of nanohole array separated by fractions of wavelength. b) Image of nanohole obtained with a conventional lens NA=1.4 where each hole can not be resolved and merge into larger formations. c) with the SOL the mean features of the cluster can be seen
Adapted from \cite{zheludev2012SOL}}
    \label{binary mask SOL nanohole}
\end{figure}
The earliest developed approach used a method touched upon earlier in the section: instead of a continuous mask (like the one that produces the Talbot ``carpet" in Fig.\ref{nanohole focus}) a binary mask SOL consisting on a series of concentric rings of varying width and diameter would ensure the accurate constructive interference of the incoming waves leading, in turn, to a sub-wavelength size hotspot where light is allowed to superoscillate\cite{zheludev2012SOL}. The reasoning behind the use of a binary mask is it guarantees the propagation of the focused light beyond the range of the evanescent waves and into the far field where purely superoscillatory effects take place, as first proposed by Berry \cite{zheludev2012SOL, Berry_2006_time}. Furthermore, with a carefully optimised binary mask SOL the throughput efficiency of the lens improves considerably, reducing potential noise effects.

The mask itself is the result of a computer algorithm aimed at optimising the individual features of these rings. This algorithm sifts through every potential construction of the SOL and assigns a weighting factor (merit function) to each possible feature and their resulting focal spot size until, combining the most efficient ones, a completely optimised focal spot, and hence SOL, is reached\cite{rogers2013optical}. As such, the concentric ring sizes and the number of them in the SOL are returned by the algorithm (Fig.\ref{binary mask SOL}a). This results in a very narrow superoscillatory spot of size 185nm where light ($\lambda=640$nm) is focused into, with very large, diffuse sidelobes on the sides as result of the superoscillatory behaviour outside the hotspot (Fig.\ref{binary mask SOL}b-c). The challenge with this, and in fact any, SOL comes with overcoming these sidelobes and obtaining an image that resolves the small central spot, while ignoring these very intense unwanted features. 

One way to do so involves only imaging objects a fraction of the size of the bright ring Fig.\ref{binary mask SOL}c). This is done in a manner similar to that of a confocal microscope: by scanning the superoscillatory spot across, e.g. a set of apertures much smaller than this unwanted bright ring (Fig.\ref{binary mask SOL}d,f), the ring will never intersect them and thus its effect is nominal on the resulting image formation (Fig.\ref{binary mask SOL}e,g) dominated from the light scattered from the central focal spot. Furthermore, a set of nanohole arrays (Fig.\ref{binary mask SOL nanohole}), separated by distances a fraction of a wavelength can too be measured with this approach and, similarly, the SOL image gives a much improved resolution than its confocal microscope counterpart of numerical aperture (NA) 1.4\cite{zheludev2012SOL}.

This firmly establishes the idea that a SOL can, under some very sensitive circumstances, image objects well below the conventional Abbe-Rayleigh diffraction limit in the far-field, Fig.\ref{binary mask SOL}h), recreating the methodology of a confocal microscope. Similarly, the SOL approach does not only improve on conventional resolution but the impact of noise is reduced. Many other superresolution techniques, particularly those in which computational methods for resolution improvement are used, are highly susceptible to noise in the image, only offering small gains in resolution when a low level of noise is present. Super-oscillatory imaging, however, is highly immune to noise, where it only results in a drop in contrast while resolution remains unchanged\cite{rogers2013optical}. It should be noted, however, this noise is different from the discussion in $\S$\ref{superoscillations and noise}, where noise in superoscillations was introduced as a destructive phase term\cite{Berry_2016noise}. Here the optical noise, instead, represents the fraction of detected photons that do not contribute constructively to the image formation and instead would typically reduce its quality.

Moreover, focusing in these type of lenses has been achieved in both achromatic (two wavelengths) and apochromatic (three wavelengths) regimes in an attempt to reduce the chromatic aberration otherwise present in SOLs due to their unavoidable sensitivity\cite{berry2019roadmap}.

\subsubsection{Dynamic SOL with spatial light modulators}

\begin{figure}
\centering
    \begin{subfigure}[t]{0.9\textwidth}
    \centering
    \includegraphics[scale=2.5]{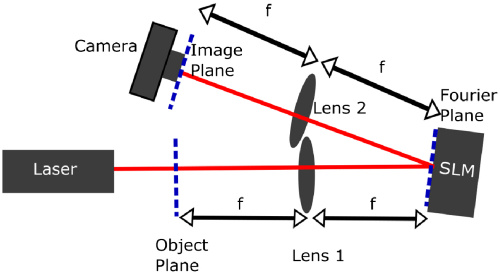}
    \end{subfigure}
    \begin{subfigure}[t]{0.9\textwidth}
    \centering
    \includegraphics[scale=0.5]{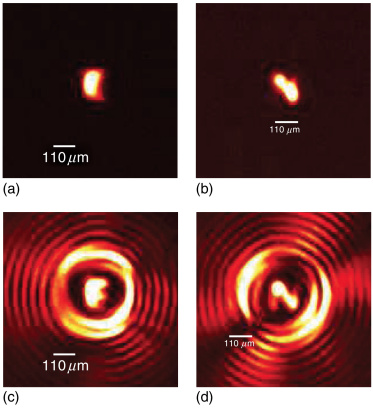}
    \end{subfigure}
    \caption{Top: Schematic of experimental setup of SLM superoscillatory imaging. Laser illuminates object with collimated light 633nm in the object plane, the wave propagates until it reaches the SLM from where it inherits the superoscillatory pattern present in its Fourier transform. Light then reflects to the camara where superoscillatory light can be seen to show the fine detail of the object.
    a-b) Microscale letters and their fine structure left unresolved by conventional diffraction limited lens. c-d) Letters can be more clearly resolved with the use of the SOL with SLM. Bright surrounding disk remains present as sidelobes are unavoidable but better resolution of smaller object is seen in hotspot.Adapted from \cite{Dong2017SLM}}
    \label{SLM}
\end{figure}
Although binary masks provides powerful and relatively simple to implement sub-wavelength hot-spot generation, it is a technique that is not all that practical. Most likely, the object to measure won't consist of apertures in an opaque metal plate but rather small irregular structures. To this end, super-oscillatory focus can also be achieved using spatial light modulators (SLM) and a conventional microscope objective that focuses a beam with a maximised amplitude and phase profile for highly sensitive, custom-fitted interference effects. The reasoning behind this alteration being that the object need not be shone with superoscillatory light to obtain the benefits of optical superoscillations. Instead, the object can be illuminated with a standard laser light beam and impose a superoscillatory pattern onto the back focal plane of it by reflecting it off SLMs\cite{Dong2017SLM}. These SLMs allow one to precisely program any desired superoscillatory pattern onto the incident collimated light where the fields of the SLMs are related to the focal plane of the objective by a Fourier transform\cite{rogers2013optical}. 

To highlight the impressive results that can be achieved with this technique, it is possible to illuminate various letters of the alphabet with a laser beam. This light will then go through the imaging system where it will be deflected off an SLM, which in turn is received at the detector as the superoscillatory pattern of the image. The improvement on a standard imaging mode is clearly visible in Fig.\ref{SLM}a-d), where the bright disk around the hotspot can be seen arising again, from the sidelobe intensity. To achieve super-oscillatory focusing in this way, the amplitude and phase pattern of the wave incident on the objective that will create the desired focal spot must be determined. The optical eigenmode method has proven to be the best suited for this purpose. It consists of probing the system with a series of masks from where the intensity and spot size matrix operators of the system are determined and optimally solved, giving always the smallest spot size possible\cite{Zheludev2011eigenmode}. One of the main benefits of introducing an SLM to a SOL is that of improving the working distance of the system: with an SLM, focal spots may be formed on the focus of a microscope objective up to around 100$\mu$m, a tenfold improvement on what a binary SOL might achieve. These and other improvements come at the expense of having to work with an overall more complex equipment compared to the binary SOL, with precise alignment of their many components required. Similarly, the chromatic aberrations any SOL is prone to are harder to remove with an SLM in place and, should they be reduced, it would further add to the complexity of the system\cite{rogers2013optical}.

\subsubsection{Limitations of SOLs}

Although superoscillations, in theory, have no physical limitations on the spot size they can create, their specific imaging implementations discussed above come with some constraints. The first and most obvious barrier is technical. Ideal SOLs of arbitrary design can be achieved sets of masks with different purposes. These masks are required to do a highly sensitive work and on top of that, must be fabricated to nanoscale accuracy. Current lens fabrication techniques are incapable of delivering such masks and real SOL masks often suffer performance issues such as pixilation\cite{rogers2013optical}. 

The second limitation is that of scattering from the sidebands into the hotspot.  As the spot size is reduced, the sideband intensity relative to the spot intensity will always increase, producing, in turn, more scattering. This effect has been studied and simulations have managed to reduce the relative energy concentration in the sidebands which, in turn, very slightly improves the resolution limit on the image (from 0.25$\lambda$ to 0.15$\lambda$)\cite{rogers2013optical}.

\subsection{Other uses for optical superoscillations}

Even though the imaging capabilities of superoscillatory lenses garner most of the attention of superoscillations in optics, some similar constructions offer other interesting qualities. One such example is that of the optical needle SOL (or ONSOL), \cite{rogers2013needle}. These ONSOLs focus light
into a sub-wavelength needle, instead of the spot associated with the standard SOLs. Besides, the needle is clearly separated from the intense sidebands, resulting in a larger field of view for the ONSOL to work in. Their construction crucially varies from the SOL in that the central area of the ONSOL blocks the incident light, forming an opaque region where the needle is formed. One of the uses of these needles is in heat-assisted magnetic recording (HAMR) in magnetic hard drives for data storage. During the magnetic writing process, components of the hard drive are locally heated leaving behind a signature. Current technology is highly inefficient at heating at the microscales most hard drives require. With the continuous improvement in disk storage over the years, this is expected to become a problem, and even more so when hard drives eventually reach the scales at which superparamagnetic effects have to be considered. Therefore the ONSOL could be a promising possibility \cite{rogers2013optical}.

\section{Conclusions}

The purpose of this review was to provide the reader with a clear understanding of the quantum mechanical framework in which superoscillations in band-limited functions arise, applying the notion of weak measurements and weak values developed by Aharonov. Despite the unintuitive idea of functions oscillating faster than their fastest bounded Fourier components, it has been shown this phenomenon is not rare and, in some cases, its effects are quite substantial. To this end, superoscillations are at the heart of Aharonov's unexpected results, explained from the perspective of large pointer shifts under a weak measurement. Similarly, the presence of superoscillatory regions demands there to be slower oscillating regions or sidebands where the vast majority of the total intensity will concentrate. This tradeoff will always be present due to the nature of superoscillations and so control over these large, diffuse sidebands proves to be key when trying to use these sensitive functions constructively.

One of the areas in which superoscillations have been extensively studied for their practicality is in optical superresolution, where arbitrarily small hotspots corresponding to superoscillatory regions can be achieved by means of a focusing device, e.g. optical masks. Superoscillations in superresolution techniques were proposed from the start due to their persistence under time evolution when used as initial data for the Schrödinger equation. This prediction has more recently been shown to allow sub-wavelength imaging of complex objects in the far field in a highly controlled manner, with experiments exploiting the optical capabilities of superoscillations by means of superoscillatory lenses (SOL). 

However, extension into imaging of more complex objects with scientific importance still remains a challenge. Furthermore, these SOLs have now become highly complex pieces of engineering and so the viable construction of such lenses will be one of the many barriers further research into the field will face. Nevertheless, further theoretical analysis will enable superoscillations to be more fully exploited in applications. This might include studying superoscillatory functions further in their own right, to gain a deeper understanding of their properties, as well as optimising the characteristics of superoscillations that are relevant to applications. All things considered, it is an exciting time in the field of superoscillations as theoretical models, driven by improving numerical methods, prove to have a growing impact on an ever-increasing number of real-world applications.

\section*{References}

\bibliographystyle{ieeetr}

\bibliography{main}

\end{document}